\documentclass[journal]{IEEEtran}
\def\FigureThreeWidth{.15}
\def\FigureThreeWidthEPS{1.1}
\usepackage{graphicx}
\usepackage{multirow}
\usepackage{amsmath}
\usepackage{amssymb}
\usepackage{float}
\usepackage[dvipsnames]{xcolor}
\usepackage{amsfonts}
\usepackage{stfloats}
\usepackage{lipsum}
\usepackage{mathtools}
\usepackage{cuted}
\usepackage{caption}
\usepackage{subcaption}
\usepackage{amsthm}
\usepackage{multicol}
\usepackage[normalem]{ulem}
\usepackage{soul}
\usepackage{datetime}
\usepackage[english]{babel}

\ifCLASSINFOpdf
\else
\fi
\begin{document}

\title{Frequency Reversal Alamouti Code-Based FBMC with Resilience to Inter-Antenna Frequency Offsets}

\author{Cheng-Yu Lin, \IEEEmembership{Student Member,~IEEE,}
        Borching~Su,~\IEEEmembership{Member,~IEEE,}
        and~Kwonhue~Choi,~\IEEEmembership{Senior Member,~IEEE}% <-this % stops a space
}
\markboth{
%To be submitted to IEEE WCL (Ver. 0.9.5, 
%1:20pm (GMT+8), September 12, 2022)
% \currenttime~(UTC)~\today).
}%
{Shell \MakeLowercase{\textit{et al.}}: Bare Demo of IEEEtran.cls for IEEE Journals}

\maketitle
\begin{abstract}
Transmit diversity schemes for filter bank multicarrier (FBMC) are known to be challenging. 
No existing schemes have considered the presence of inter-antenna frequency offset (IAFO), which will result in performance degradation. 
In this letter, a new transmit scheme based on the frequency reversal Alamouti code (FRAC)-based structure to address the issue of IAFO is proposed and is proven to inherently cancel the inter-antenna inter-carrier interference (ICI) while preserving spatial diversity. 
Moreover, the proposed FRAC structure is applicable in frequency-selective channels.
Numerical results show that the proposed scheme undergoes negligible bit error rate (BER) degradation even with considerable IAFOs.
%\footnote{This work is supported jointly by the National Research Foundation of Korea (NRF) grant funded by the Korean government (MSIT) under Grant 2021R1A2C1010370, and by the Ministry of Science and Technology (MOST) of Taiwan under grant number  MOST 110-2221-E-002-074.
%\emph{(Corresponding author: Kwonhue Choi.)}  Kwonhue Choi is with the Department of Information and Communication Engineering, Yeungnam University, Gyeongsan 38541, South Korea (e-mail: gonew@yu.ac.kr). 
%Cheng-Yu Lin and Borching Su are with the Graduate Insitute of Communication Engineering, National Taiwan University, Taipei 10617, Taiwan.}
\end{abstract}


\IEEEpeerreviewmaketitle

\section{Introduction}
Filter bank multicarrier (FBMC) is a promising candidate modulation of beyond-5G systems. 
However, the implementation of the MIMO technique on filter bank multicarrier (FBMC) is always a challenging issue due to the presence of the interference terms, which do not exist in some other multicarrier schemes. 
Some references have proposed solutions to tackle this issue.
% Some references have proposed the approachs of Alamouti coding on FBMC.
For instance, the typical method for Alamouti coded FBMC was proposed in \cite{Renfors 2010}, and a block-wise Alamouti coding scheme is constructed. Moreover, the time-reversal (TR) block structure shows outstanding performance when the system experience a stationary channel. Another important method was proposed in \cite{Nam 2016}, which separates the QAM symbols into odd-numbered subcarrier symbols and even-numbered subcarrier symbols and then applies two prototype filters.

However, in some scenarios of cell systems, e.g., distributed antenna systems (DAS) \cite{Moerman 2022}, and coordinated multipoint (CoMP) \cite{Elhattab 2022}, distributed TX antennas have different frequency offsets because the individual TX antenna experiences different factors such as different RF chains or different Doppler frequencies. %are assigned with different frequencies and involved for the same user equipment (UE). 
As a result, the inter-antenna frequency offset (IAFO) is  inevitable.  
Unfortunately, existing methods do not consider the IAFO. 
Thus, they cause inter-antenna inter-carrier interference (ICI) and lead to bit error rate (BER) performance degradation.

In this letter, considering a situation where the signals from the two transmitters have different frequency offsets, we adopt the frequency reversal Alamouti code-based-FBMC (FRAC-FBMC) scheme proposed in \cite{Na-Choi-2016} and \cite{Na-Choi-2017} and extend it to multiple subblocks. 
The contribution of this letter is three-fold:
\begin{itemize}
\item A new FRAC-FBMC transmit and processing method is proposed specifically for tackling the effect of IAFO. The multi-subblock design makes the scheme applicable even in  frequency-selective channels.
\item
The proposed FRAC-FBMC is shown, through rigorous mathematical derivations, to cancel the inter-antenna ICI that caused by IAFO, without any cancellation algorithm.
%\item
%The effect of half-subblock size and its optimization under frequency-selective channels are provided. 
\item
Intensive comparisons with the state-of-the-art Alamouti coded FBMCs are provided to show the superiority of the proposed FRAC-FBMC.
\end{itemize}

\begin{figure*}[ht]
        \centering
        \includegraphics[width=1\textwidth,clip]{Figure_1.pdf}
       
        %\captionsetup{labelformat=empty}
        \caption{Alamouti coded symbol mapping of the proposed frequency reversal scheme ($L_n=1$).}
\label{fig:Frequency-Reversal-Alamouti-Coded}
    \end{figure*}
    
\section{IAFO-Free FRAC-FBMC}
\subsection{Transmit Signal structure}
We consider an $N$-subcarrier FBMC-OQAM system with two transmit antennas and one receive antenna.
Assume the two antennas, namely antenna $A$ and antenna $B$, transmit the FBMC-OQAM-modulated signals as
\begin{equation}
s_a(t)=\sum_{l=1}^{N}\sum_{m=-\infty}^{\infty}a_{l,m}\zeta^{(a)}_{l,m}p\left(t-\frac{mT}{2}\right)e^{j\frac{2\pi lt}{T}},
\label{eq:s_a}\end{equation}
\begin{equation}
s_b(t)=\sum_{l=1}^{N}\sum_{m=-\infty}^{\infty}b_{l,m}\zeta^{(b)}_{l,m}p\left(t-\frac{mT}{2}\right)e^{j\frac{2\pi lt}{T}}
\label{eq:s_b}\end{equation}
where $T$ is the FBMC symbol duration, $l$ is the subcarrier index, $m$ is the half-symbol (time) index, and $p(t)$ is the pulse shaping prototype filter.
The information-bearing symbols, denoted $a_{l,m}$ and $b_{l,m}$ for antennas $A$ and $B$, respectively, are real-valued. 
The terms $\zeta_{l,m}^{(a)}$ and $\zeta_{l,m}^{(b)}$ are phase shift terms whose properties will be described later in this subsection. 

The $N$ subcarriers are partitioned into $\mathcal{K}$ subblocks of size $N_F$, i.e., $\mathcal{K}=N/N_F$.
The subcarriers in the $i$th subblock are indexed from $iN_F+1$ to $(i+1)N_F$ for all $i=0, 1, ...,\mathcal{K}-1$.
For the $i$th subblock in the $n$th FBMC subsymbol, we divide it into two halves and exploit $L_n$ null subcarriers to the left of each half-subblock, i.e., $a_{k+iN_F,n}=b_{k+iN_F,n}=0$ for all $k=1,..., L_n$ and $k=N_F/2+1,..., N_F/2+L_n$.
The rest of the subcarriers in the subblock are encoded by $N_F-2L_n$ real-valued information-bearing symbols $x_{k,n}^{(i)}$ and $y_{k,n}^{(i)}$, $k=L_n+1,...,N_F/2$,
according to the following frequency-reversal Alamouti coding:
 \begin{equation}
    \begin{pmatrix}
    a_{k+iN_F,n}&a_{(i+1)N_F-k+1+L_n,n}\\ b_{k+iN_F,n}&b_{(i+1)N_F-k+1+L_n,n}
    \end{pmatrix}=\begin{pmatrix}x^{(i)}_{k,n}&-y^{(i)}_{k,n}\\ y^{(i)}_{k,n}&x^{(i)}_{k,n}\end{pmatrix}
    \label{eq:ala}%\tag{3'}
\end{equation} 
for all $i=0,…, \mathcal{K}-1$. 
Fig. \ref{fig:Frequency-Reversal-Alamouti-Coded} illustrates the idea of the FRAC symbol mapping described above for the special case $L_n=1$.  
The phase shift terms $\zeta_{l,m}^{(a)}$ and $\zeta_{l,m}^{(b)}$ alternate between $1$ (or $-1$) and $j$ (or $-j$) in both of time and frequency axes, and follow the rules similar to those of [\cite{Na-Choi-2017}, (25) and (26)], i.e., 
$\zeta^{(a)}_{(i+1)N_F-l+L_n + 1, m} = \chi \zeta^{*(b)}_{l+iN_F, m}$ and $\zeta^{(b)}_{(i+1)N_F-l+L_n + 1, m} = \chi \zeta^{*(a)}_{l+iN_F, m}$ for all $i$ and $l$, $1+L_n\leq l\leq N_F/2$ and $\chi=\pm j$ (or $\pm 1$), where the superscript $(\cdot)^*$ denotes the complex conjugate.

\subsection{Received Signal and Demodulation}
We assume there exist carrier frequency offsets among the two transmitters and the receiver. 
Denote $f_a$ and $f_b$ as the carrier frequency offsets of transmitters $A$ and $B$, respectively, relative to the carrier frequency of the receiver.
Let the transmit signals from transmitters $A$ and $B$ pass through frequency-selective channels $h_a(t)$ and $h_b(t)$, respectively. 
Then, the received signal can be expressed as follows:

{\setlength\abovedisplayskip{-7pt}
\setlength\belowdisplayskip{2pt} 
\begin{equation}
        \begin{aligned}r(t)&=e^{j2\pi f_at}\int_{0}^{\Delta} h_a(\tau) s_a(t-\tau)d\tau\\
        &\quad+e^{j2\pi f_bt}\int_{0}^{\Delta} h_b(\tau)s_b(t-\tau)d\tau+n(t)\\
        &=e^{j2\pi f_at}\sum_{i=0}^{\mathcal{K}-1}\sum_{l=1}^{N_F}\sum_{m=-\infty}^{\infty}a_{l+iN_F,m}\zeta^{(a)}_{l,m}\\
        & \quad \cdot \int_{0}^{\Delta} h_a(\tau)p\left(t-\tau-\frac{mT}{2}\right)e^{j\frac{2\pi (l+iN_F)(t-\tau)}{T}}d\tau\\
        &\quad+e^{j2\pi f_bt }\sum_{i=0}^{\mathcal{K}-1}\sum_{l=1}^{N_F}\sum_{m=-\infty}^{\infty}b_{l+iN_F,m}\zeta^{(b)}_{l,m}\\
        & \quad \cdot \int_{0}^{\Delta} h_b(\tau)p\left(t-\tau-\frac{mT}{2}\right)e^{j\frac{2\pi (l+iN_F)(t-\tau)}{T}}d\tau\\
        &\quad+n(t)
    \end{aligned}\label{eq:r_t_mp}\end{equation}}
    
\noindent where $\Delta$ is the maximum delay spread of the channels, and $n(t)$ is an additive complex white Gaussian noise with power spectral density $N_0/2$.

We assume that $f_a$, $f_b$, $h_a(t)$, and $h_b(t)$ are known to the receiver.
Next, we consider the $k$th subcarrier of the $q$th subblock in the $n$th FBMC subsymbol, with the received samples $r_{k,n}^{(a,q)}$, $r_{k,n}^{(b,q)}$ being calculated by FBMC demodulation with respect to frequency offsets $f_a$, $f_b$, and phase shift terms $\zeta_{l,m}^{(a)}$, $\zeta_{l,m}^{(b)}$, respectively:

  {\setlength\abovedisplayskip{-5pt}
\setlength\belowdisplayskip{0pt} 
 \begin{equation}\begin{aligned}
        r^{(a,q)}_{k,n}=\int_{-\infty}^{\infty}e^{-j2\pi f_at}r(t)\zeta^{*{(a)}}_{k,n}p\left(t-\frac{nT}{2}\right)e^{-j\frac{2\pi (k+qN_F)t}{T}}dt
        \label{eq:ra_k_n_flat}\end{aligned}\end{equation} 
        \begin{equation}\begin{aligned}
        r^{(b,q)}_{k,n}=\int_{-\infty}^{\infty}e^{-j2\pi f_bt}r(t)\zeta^{*{(b)}}_{k,n}p\left(t-\frac{nT}{2}\right)e^{-j\frac{2\pi (k+qN_F)t}{T}}dt
        \label{eq:rb_k_n_flat}\end{aligned}\end{equation}}

\noindent for $k=1, 2,..., N_F$, $q=0, 1,..., \mathcal{K}-1$.
We define $H_{a}^{(q, k)}$ (or $H_{b}^{(q, k)}$) as the channel response of $h_a(t)$ (or $h_b(t)$) at the $k$th subcarrier of the $q$th subblock, i.e., \begin{equation}H_{\star}^{(q, k)} = \int_{0}^{\Delta} h_{\star}(t) e^{-j\frac{2\pi (k+qN_F)t}{T}} dt\end{equation}
where $\star$ can be $a$ or $b$. 
We assume that $\Delta$ is sufficiently small so that a quasi-static fading assumption holds, i.e., $H_a^{(q,k)}$ and $H_b^{(q,k)}$ can be regarded as constant within the $q$th subblock, and we define $H_a^{(q)} \approx H_a^{(q,k)}$ and $H_b^{(q)} \approx H_b^{(q,k)}$. 
Next, the decision variables for data symbols $x^{(q)}_{k,n}$ and $y^{(q)}_{k,n}$  with $k=L_n+1, L_n+2, ..., N_F/2$, and $q=0, 1, ..., \mathcal{K}-1$, are obtained by combining the two received samples as follows:
\begin{equation}
    d_{k,n}^{(x,q)}=\Re\left[H^{*(q)}_{a}r^{(a,q)}_{k,n}+H^{(q)}_{b}r^{*(b,q)}_{N_F-k+1{+L_n},n}\right],
\label{eq:dx_k_n}\end{equation}
\begin{equation}
    d_{k,n}^{(y,q)}=\Re\left[H^{*(q)}_{b}r^{(b,q)}_{k,n}-H^{(q)}_{a}r^{*(a,q)}_{N_F-k+1+L_n,n}\right]
\label{eq:dy_k_n}\end{equation}
where $\Re[x]$ denotes the real part of $x$. In the following developments, for simplicity, we only consider the $0$th subblock (i.e., $q=0$), noting that cases in other subblocks can be easily extended and will obtain identical results. 
In this regard, we let $H_a\triangleq H_a^{(0)}$, $H_b\triangleq H_b^{(0)}$, $d_{k,n}^{(x)}\triangleq d_{k,n}^{(x,0)}$, $d_{k,n}^{(y)}\triangleq d_{k,n}^{(y,0)}$, $x_{k,n}\triangleq x^{(0)}_{k,n}$, $y_{k,n}\triangleq y^{(0)}_{k,n}$, $r_{k,n}^{(a)}\triangleq r_{k,n}^{(a,0)}$, and $r_{k,n}^{(b)}\triangleq r_{k,n}^{(b,0)}$.
We will show that $d_{k,n}^{(x)}$ (or $d_{k,n}^{(y)}$) and $x_{k,n}$ (or $y_{k,n}$) have a relationship of diversity gain, i.e., 
\begin{equation}d^{(x)}_{k,n}=(|H_a|^2+|H_b|^2)x_{k,n}+n_{k,n}^{(x)},\label{eq:d_k_n_x}\end{equation}
\begin{equation}
d^{(y)}_{k,n}=(|H_a|^2+|H_b|^2)y_{k,n}+n_{k,n}^{(y)}\label{eq:d_k_n_y}\end{equation}
\noindent where operator $|\cdot|$ denotes the absolute value. 
To see the validity of \eqref{eq:d_k_n_x}, we substitute (\ref{eq:s_a}) and (\ref{eq:s_b}) into (\ref{eq:r_t_mp}) and then into (\ref{eq:ra_k_n_flat}). 
After performing some variable substitutions, we write $r^{(a)}_{k,n}$ as follows:
{\setlength\abovedisplayskip{0pt}
\setlength\belowdisplayskip{0pt} 
\begin{equation}
\begin{aligned}r^{(a)}_{k,n}
&=H_a \sum_{l=-L}^{L}\sum_{m=-M}^{M}a_{l+k,m+n}\zeta^{(a)}_{l+k,m+n}\zeta^{*(a)}_{k,n}F^{(0)}_{l,m,n}\\
&\quad+H_b\sum_{l=-L}^{L}\sum_{m=-M}^{M}b_{l+k,m+n}\zeta^{(b)}_{l+k,m+n}\zeta^{*(a)}_{k,n}F_{l,m,n}^{(f_b-f_a)}\\&\quad+w_{k,n}
\end{aligned}
\label{eq:rra_k_n_2}\end{equation}}

\noindent where $F^{(\cdot)}_{l,m,n}$ is defined as% and $F_{l,m, n}^{(f_a,f_b)}$ as
\newcommand{\Deltaf}{\Delta_f}
\begin{equation}
F_{l,m,n}^{(\Deltaf)} = e^{j\pi n(l+\Deltaf T)}\int_{-\infty}^{\infty} p\left(t-\frac{mT}{2}\right)p(t) 
e^{j\frac{2\pi l t}{T}}
e^{j{2\pi \Deltaf t}}
dt,
\end{equation}
$w_{k,n}$ is a zero-mean complex Gaussian noise term, and $L$ and $M$ are small positive integers determined by the localization of the pulse $p(t)$ in the time-domain and frequency-domain, respectively \cite{Bellanger 2010}. 
Note that $e^{j2\pi \Deltaf t}$ with $\Deltaf = f_b-f_a$ represents the residual carrier term due to IAFO. 
We assume that $L_n$ is chosen sufficiently large such that $L_n\geq L$.
Noting that $r_{k,n}^{(b)}$ as in \eqref{eq:rb_k_n_flat} can be treated similarly, we express $r_{k,n}^{(a)}$ and $r_{N_F-k+1+L_n,n}^{(b)}$, $L_n+1\leq k\leq N_F/2$, as:
\begin{eqnarray}
r_{k,n}^{(a)}
&=&H_a\underset{\triangleq U}{\underline{\sum_{l=-L}^{L}\sum_{m=-M}^{M}a_{l+k,m+n}F_{l,m(a\rightarrow a)}^{(k,n)}}}\nonumber\\
&&+H_b\underset{\triangleq V}{\underline{\sum_{l=-L}^{L}\sum_{m=-M}^{M}b_{l+k,m+n}F_{l,m(b\rightarrow a)}^{(k,n)}}}
+w_{k,n}\nonumber\\
&=&H_aU+H_bV+w_{k,n},
\label{eq:rra_k_n_3}\end{eqnarray}
\begin{equation}\begin{aligned}
        &r^{(b)}_{N_F-k+1{+L_n},n}\\
        &=H_a \underset{\triangleq W}{\underline{\sum_{l=-L}^{L}\sum_{m=-M}^{M}a_{l+N_F-k+1{+L_n},m+n}
        F_{l,m(a\rightarrow b)}^{(N_F-k+1{+L_n},n)}}}\\
        &\quad+H_b\underset{\triangleq Z}{\underline{\sum_{l=-L}^{L}\sum_{m=-M}^{M}b_{l+N_F-k+1{+L_n},m+n}
         F_{l,m(b\rightarrow b)}^{(N_F-k+1{+L_n},n)}}}\\
        &\quad+w_{N_F-k+1{+L_n},n}%\\
        =H_aW+H_bZ+w_{N_F-k+1{+L_n},n}
    \label{eq:rrb_k_n}
    \end{aligned}\end{equation}
    with the interference coefficients in (\ref{eq:rra_k_n_3}) and  (\ref{eq:rrb_k_n}) defined as:
    \begin{eqnarray}
         F_{l,m(a\rightarrow a)}^{(k,n)}&\triangleq& \zeta^{(a)}_{l+k,m+n}\zeta^{*(a)}_{k,n}F_{l,m,n}^{(0)},\\ 
         F_{l,m(b\rightarrow a)}^{(k,n)}&\triangleq& \zeta^{(b)}_{l+k,m+n}\zeta^{*(a)}_{k,n}F_{l,m,n}^{(f_b-f_a)},\\
         F_{l,m(a\rightarrow b)}^{(k,n)}&\triangleq& \zeta^{(a)}_{l+k,m+n}\zeta^{*(b)}_{k,n}F_{l,m,n}^{(f_a-f_b)},\\
         F_{l,m(b\rightarrow b)}^{(k,n)}&\triangleq& \zeta^{(b)}_{l+k,m+n}\zeta^{*(b)}_{k,n}F_{l,m,n}^{(0)}.
    \end{eqnarray}
Note that $V$ in (\ref{eq:rra_k_n_3}) and $W$ in (\ref{eq:rrb_k_n}) contain the inter-antenna ICI caused by IAFO. 
Now, substituting (\ref{eq:rra_k_n_3}) and (\ref{eq:rrb_k_n}) into (\ref{eq:dx_k_n}), we have
\begin{equation}
\setlength{\abovedisplayskip}{5pt}
\setlength{\belowdisplayskip}{5pt}
    \begin{aligned}&d_{k,n}^{(x)}%\\&
    =    \Re[H_a^*(H_aU+H_bV)+H_b(H_aW+H_bZ)^*]+n_{k,n}^{(x)}\\
    &=|H_a|^2\Re\left[U\right]+|H_b|^2\Re\left[Z^*\right]+\Re\left[H_a^*H_b(V+W^*)\right]+n_{k,n}^{(x)}.\end{aligned}
    \label{eq:dx_k_n_2}
\end{equation}
Due to the property of $p(t)$ and the alternating phase shift terms, we can verify the following property:
    \begin{equation}
        \Re\left[ F_{l,m(a\rightarrow a)}^{(k,n)}\right]=\Re\left[ F_{l,m(b\rightarrow b)}^{(k,n)}\right]=\delta_l \delta_m
    \label{property1}\end{equation}
where $\delta_l$ denotes the discrete-time delta function. 
Using \eqref{eq:ala} and \eqref{property1}, it is not difficult to verify that
\begin{equation}
  \label{eq:Ru_Rz_x_k_n}  \Re[U] = \Re[Z^*]= x_{k,n} 
\end{equation}with $L_n+1\leq k\leq N_F/2$.
As for the values of $W$ and $V$ which contain the IAFO terms $F_{l,m(a\rightarrow b)}^{(N_F-k+1{+L_n},n)}$ and $F_{l,m(b\rightarrow a)}^{(k,n)}$, we can show that they are self-canceled, i.e., $W^*=-V$ by performing a variable substitution as follows:
\begin{equation}\begin{aligned}W^*&=\Biggl[\sum_{l=-L}^{L}\sum_{m=-M}^{M}a_{l+N_F-k+1{+L_n},m+n}\zeta^{(a)}_{l+N_F-k+1{+L_n},m+n}\\
&\quad \cdot \zeta^{*(b)}_{N_F-k+1{+L_n},n} F_{l,m,n}^{(f_a-f_b)}\Biggr]^* \\
&=\sum_{l=-L}^{L}\sum_{m=-M}^{M}a_{N_F-k+1-l{+L_n},m+n}\zeta^{(b)}_{l+k,m+n}\zeta^{*(a)}_{k,n}F_{l,m,n}^{(f_b-f_a)}\\
&=\sum_{l=-L}^{L}\sum_{m=-M}^{M}(-b_{l+k,m+n})F_{l,m(b\rightarrow a)}^{(k,n)}\\
&=-V, \end{aligned}\label{eq:W*}\end{equation}
where, in the third equality, we had used the fact that $L_n\geq L$ and $a_{N_F-k+1{+L_n},n}=-b_{k,n}$ for any $k$ with $1-L+L_n\leq k\leq N_F/2+L$, which is justified by \eqref{eq:ala}. 
Finally, by substituting \eqref{eq:Ru_Rz_x_k_n} and (\ref{eq:W*}) into  (\ref{eq:dx_k_n_2}), the achieved diversity gain without inter-antenna interference as shown in (\ref{eq:d_k_n_x}) is verified. 
Since (\ref{eq:d_k_n_y}) can also be verified using a similar approach, the inter-antenna ICI-free characteristics of the proposed scheme are proven.
       \\
  
\section {Simulation Result}
\begin{figure*}[htb]
    \begin{subfigure}[b]{.32\textwidth}
        \centering
        \includegraphics[width=0.94\textwidth]{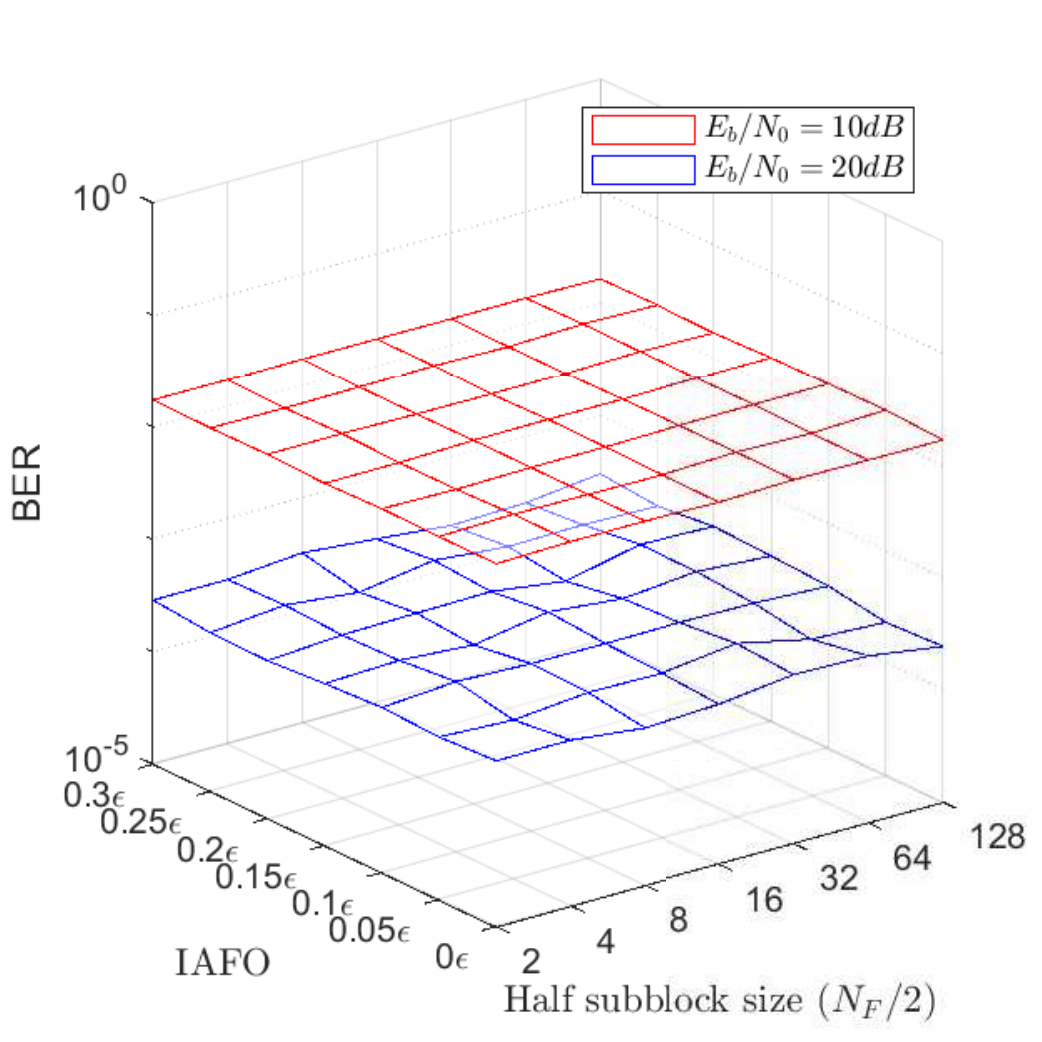}
        \subcaption{Flat fading}\label{fig:3D_BER_flat}
    \end{subfigure}
    \hfill
    \begin{subfigure}[b]{.32\textwidth}
        \centering
        \includegraphics[width=0.94\textwidth]{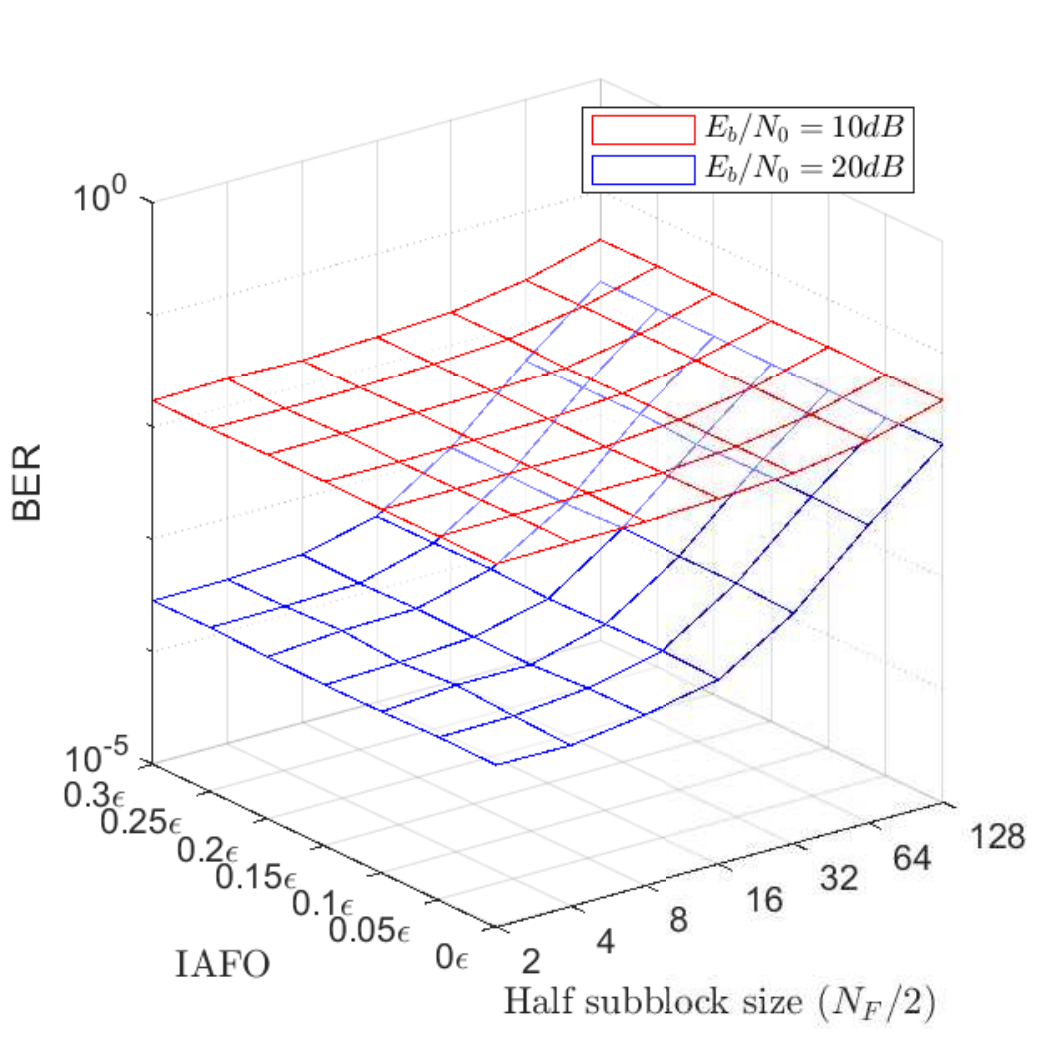}        \subcaption{ITU-R-PA}\label{fig:3D_BER_PA}
    \end{subfigure}
    \hfill
    \begin{subfigure}[b]{.32\textwidth}
        \centering
        \includegraphics[width=0.94\textwidth]{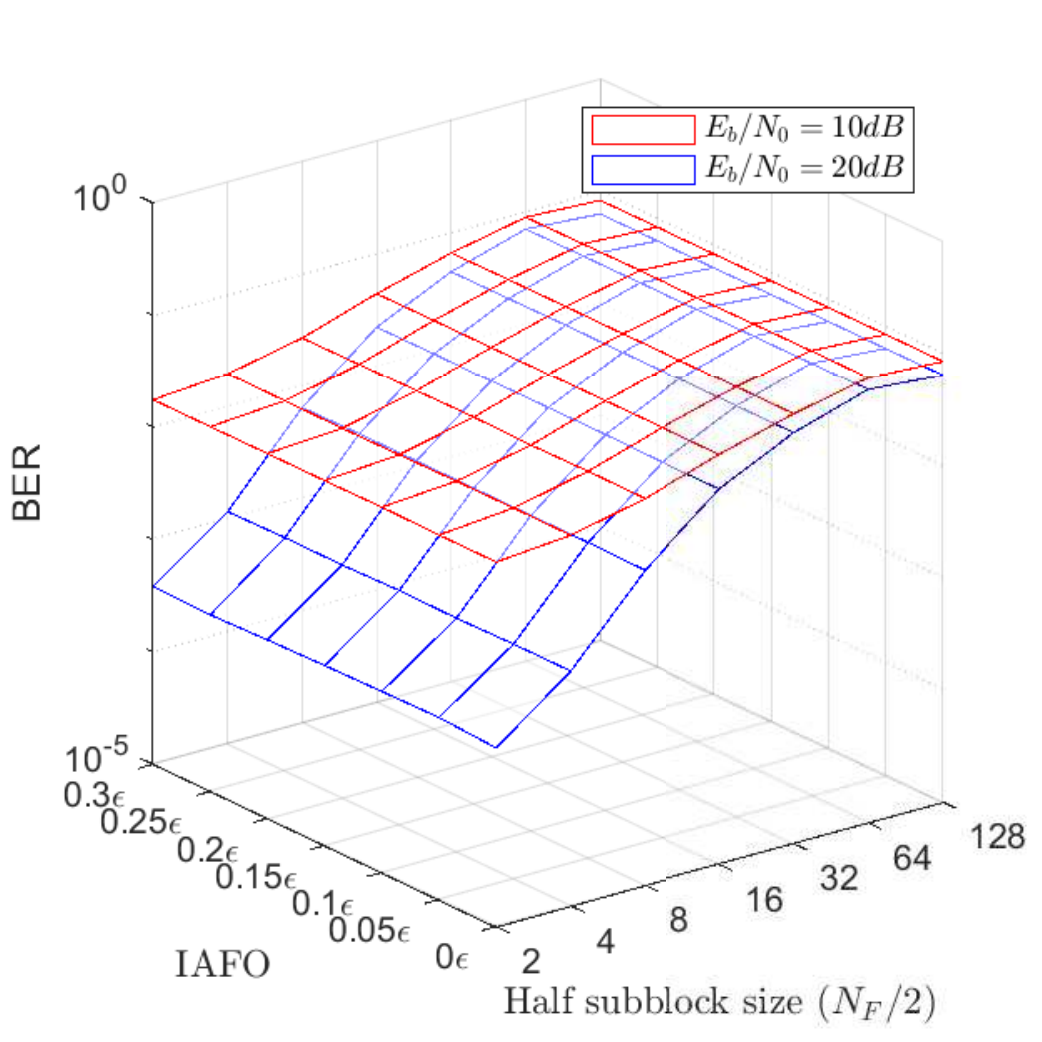}
        \subcaption{ITU-R-VA}\label{fig:3D_BER_VA}
    \end{subfigure}  
    \caption{BER curves with various half-subblock size}
    \label{fig:3D_plot}
\end{figure*}
In the simulation of this work, we use the PHYDYAS prototype filter in \cite{Bellanger 2010} with an overlapping factor {$\lambda=4$, and the corresponding localization factors $L=1$ and $M=4$.}

\subsection{BER versus different IAFOs and half-subblock size}
In Figs. \ref{fig:3D_BER_flat}--\ref{fig:3D_BER_VA}, we investigate the dependency of the proposed scheme on IAFO and simulate its BER performance according to different half-subblock sizes $N_F/2$ and different IAFOs under flat fading channel, ITU-R-Pedestrian A (ITU-R-PA), and ITU-R-Vehicular A (ITU-R-VA) multipath channels \cite{ITU-R-M.1225}. 
In this section, we set $N=256$, $L_n=1$, and the subcarrier spacing $\epsilon$ $ (=1/T)$ is set to 15kHz. 
At each simulated BER point, a total of $4\times 10^{4}$ Monte Carlo trials have been performed. 
Figs. \ref{fig:3D_BER_flat}--\ref{fig:3D_BER_VA} reveal that even with nonzero IAFOs, the proposed scheme maintains the identical BER to that with zero IAFO.
Fig. \ref{fig:3D_BER_flat} corresponds to the case of flat fading, with negligible change in the BER curves along the $N_F/2$-axis have  
negligible change.  
On the other hand, in Figs. \ref{fig:3D_BER_PA}--\ref{fig:3D_BER_VA}, with the channel selectivity increasing, choosing a large half-subblock size tends to violate the quasi-static fading assumption. 
Thus, the inter-antenna ICI can not be canceled, resulting in the degradation of BER performance. 

In practical applications, the choice of $N_F/2$ will be a compromise between bandwidth efficiency and BER degradation: choosing a large subblock size $N_F$ implies a saving in bandwidth efficiency but at the expense of a potential BER performance degradation, especially when the channel frequency selectivity is severe.
Therefore, in Figs. \ref{fig:3D_BER_PA} and \ref{fig:3D_BER_VA}, where the subcarrier spacing is $\epsilon=15$kHz, we recommend to choose $N_F/2$ to be $8$ and $4$, respectively.
\subsection{Performance comparison with the other schemes}
    \begin{figure}[h]
    \hspace*{\fill}
    \begin{subfigure}[b]{\FigureThreeWidth\textwidth}
        \centering
        \includegraphics[width=\FigureThreeWidthEPS\textwidth]{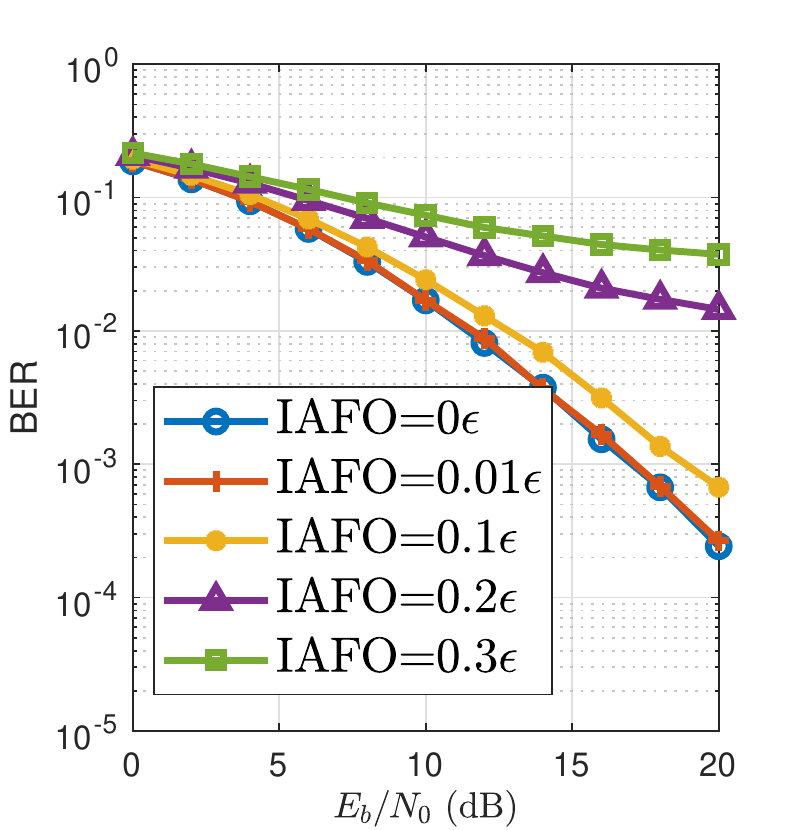}
        \subcaption{TR-FBMC}\label{fig:BER_Curves_of_TR_FBMC_flat}
    \end{subfigure}
    \hfill
    \begin{subfigure}[b]{\FigureThreeWidth\textwidth}
        \centering
        \includegraphics[width=\FigureThreeWidthEPS\textwidth]{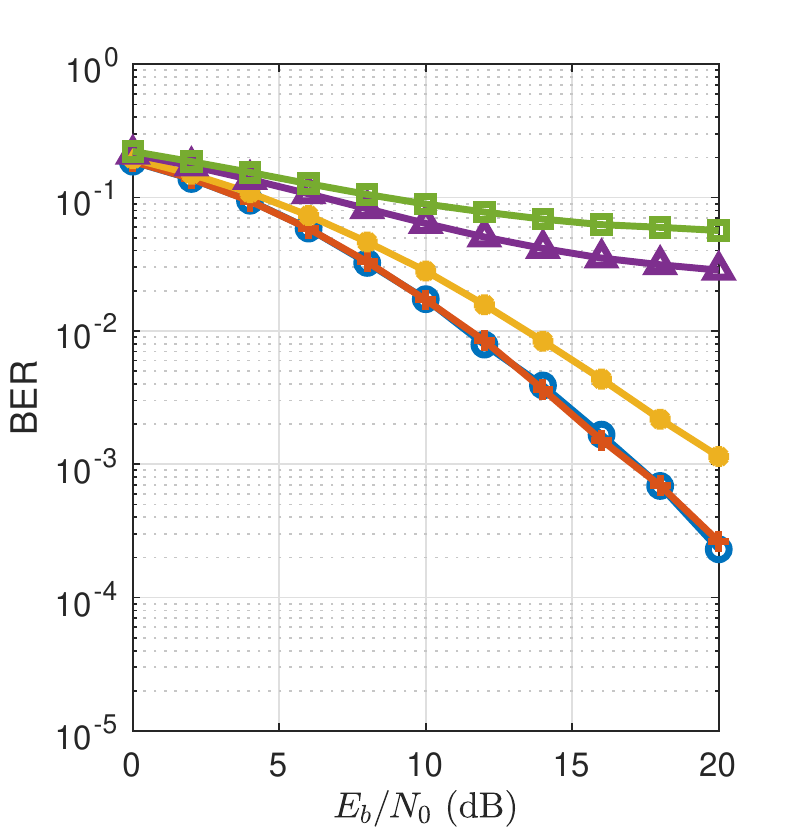}
        \subcaption{FBMC-QAM}\label{fig:BER_Curves_for_FBMC_QAM_flat}
    \end{subfigure}
    \hfill
    \begin{subfigure}[b]{\FigureThreeWidth\textwidth}
        \centering
        \includegraphics[width=\FigureThreeWidthEPS\textwidth]{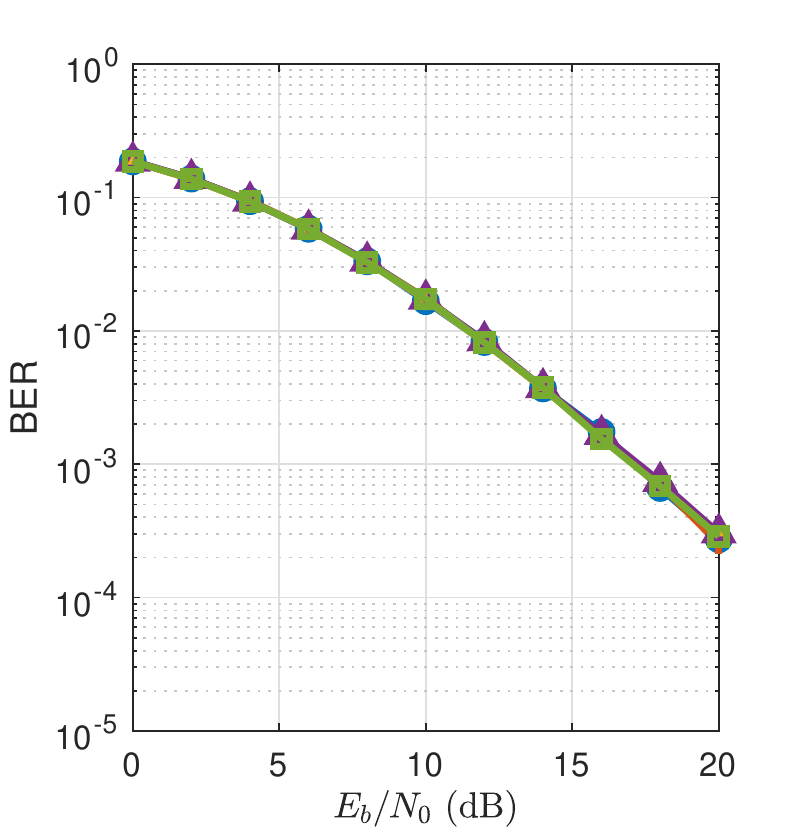}
        \subcaption{Proposed method}\label{fig:BER_Curves_of_FRAC_FBMC_flat}
    \end{subfigure} 
    \hspace*{\fill} \\
    \hspace*{\fill}
    \begin{subfigure}[b]{\FigureThreeWidth\textwidth}
        \centering
        \includegraphics[width=\FigureThreeWidthEPS\textwidth]{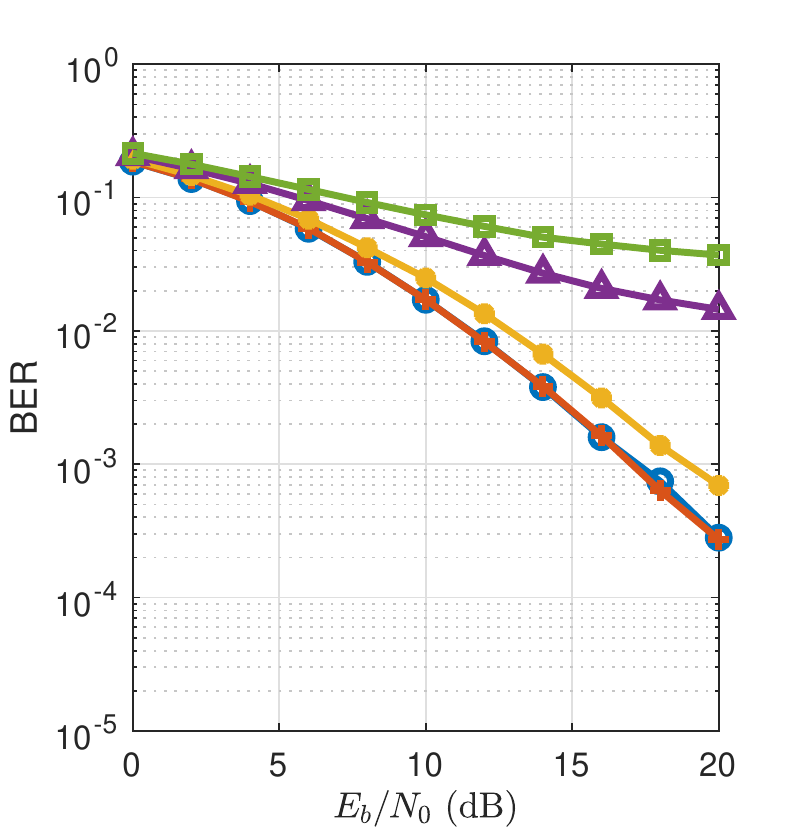}
        \subcaption{TR-FBMC}\label{fig:BER_Curves_of_TR_FBMC_PA}
    \end{subfigure}
    \hfill
    \begin{subfigure}[b]{\FigureThreeWidth\textwidth}
        \centering
        \includegraphics[width=\FigureThreeWidthEPS\textwidth]{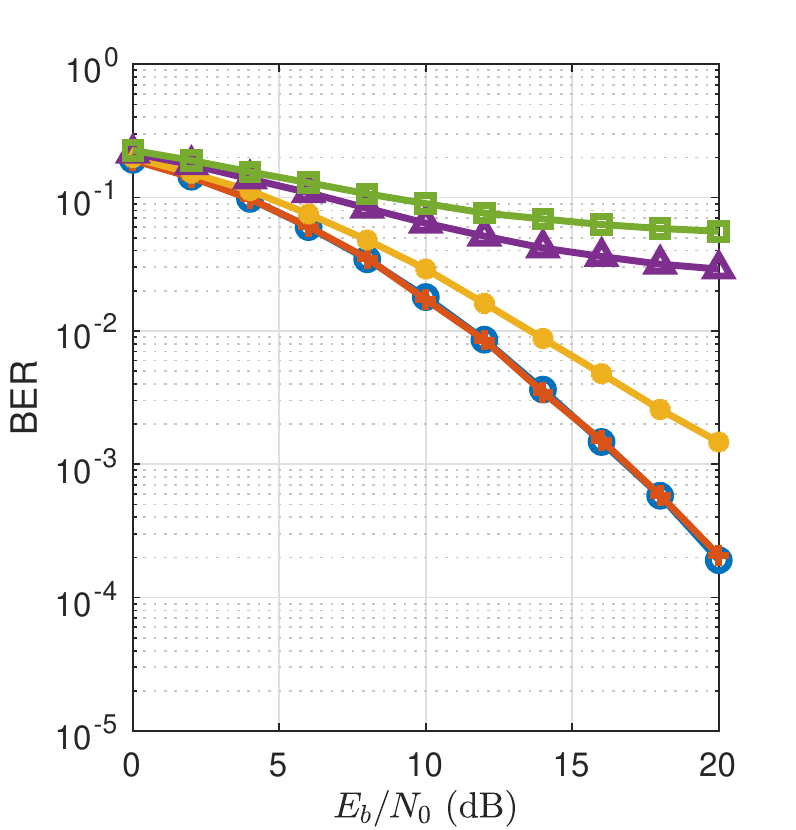}
        \subcaption{FBMC-QAM}\label{fig:BER_Curves_for_FBMC_QAM_PA}
    \end{subfigure}
    \hfill
    \begin{subfigure}[b]{\FigureThreeWidth\textwidth}
        \centering
        \includegraphics[width=\FigureThreeWidthEPS\textwidth]{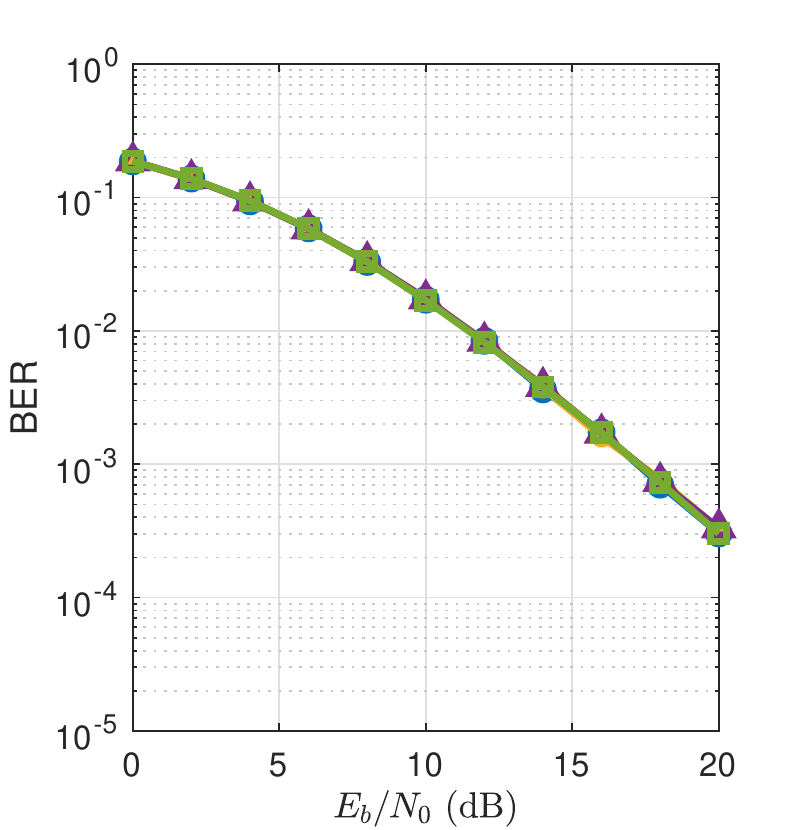}
        \subcaption{Proposed method}\label{fig:BER_Curves_of_FRAC_FBMC_PA}
    \end{subfigure}  
    \hspace*{\fill} \\
    \hspace*{\fill}
    \begin{subfigure}[b]{\FigureThreeWidth\textwidth}
        \centering
        \includegraphics[width=\FigureThreeWidthEPS\textwidth]{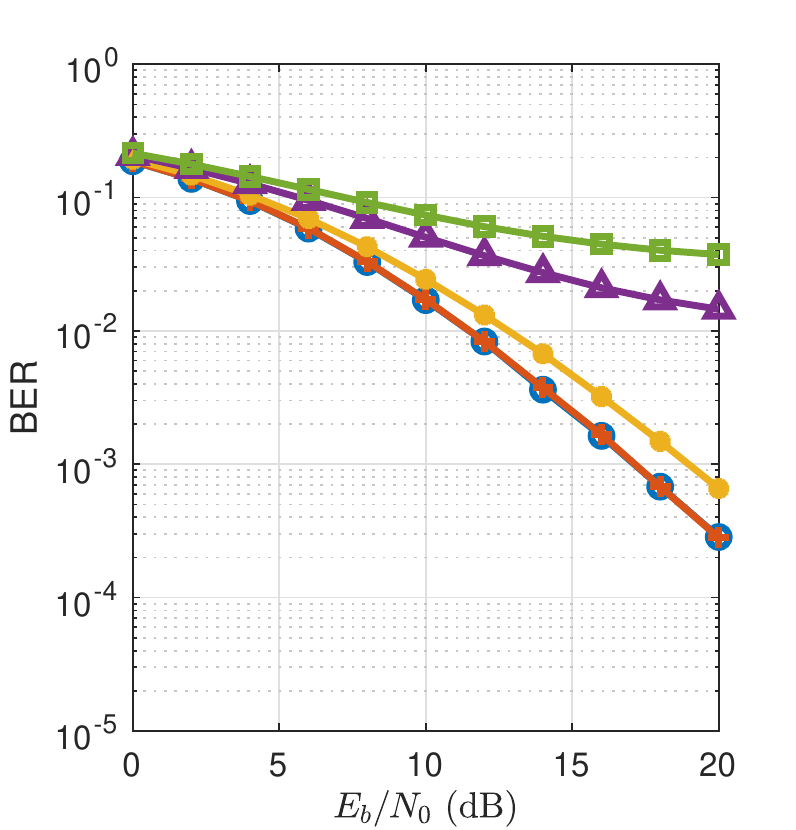}
        \subcaption{TR-FBMC}\label{fig:BER_Curves_of_TR_FBMC_VA}
    \end{subfigure}
    \hfill
    \begin{subfigure}[b]{\FigureThreeWidth\textwidth}
        \centering
        \includegraphics[width=\FigureThreeWidthEPS\textwidth]{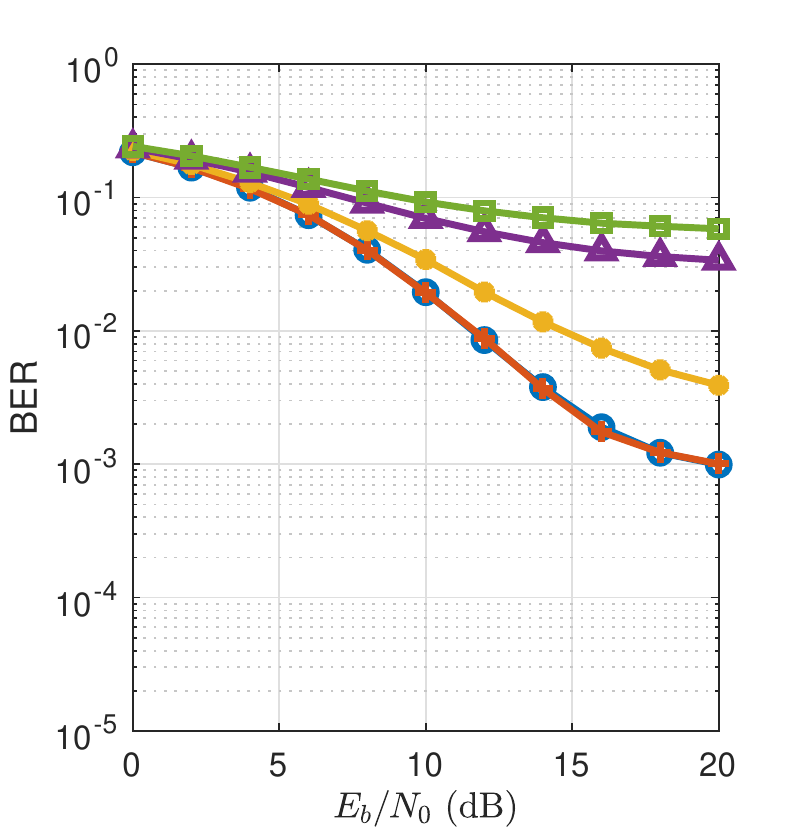}
        \subcaption{FBMC-QAM}\label{fig:BER_Curves_for_FBMC_QAM_VA}
    \end{subfigure}
    \hfill
    \begin{subfigure}[b]{\FigureThreeWidth\textwidth}
        \centering
        \includegraphics[width=\FigureThreeWidthEPS\textwidth]{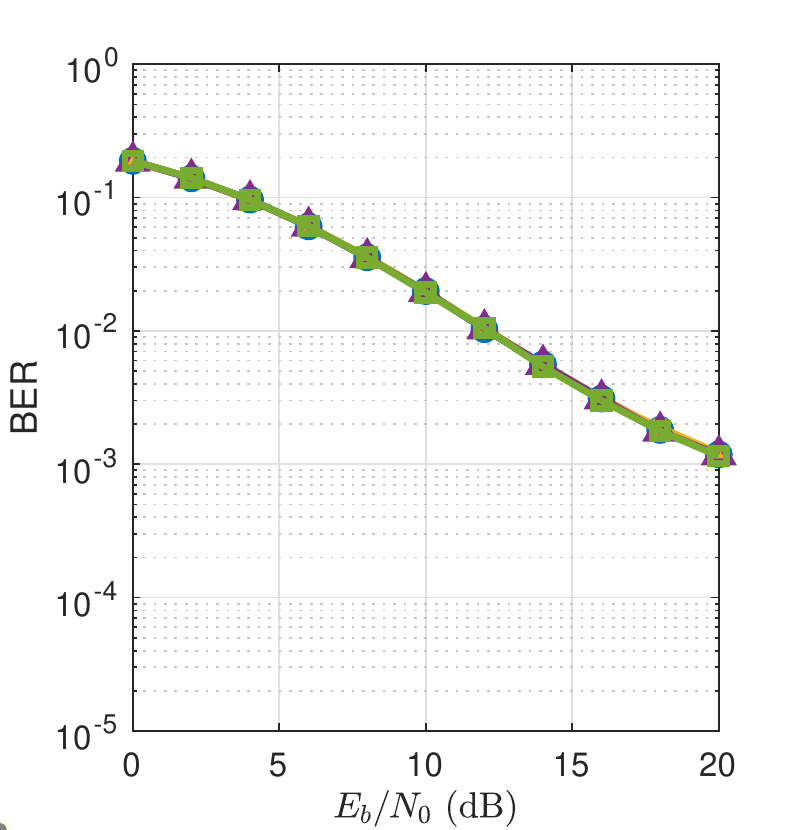}
        \subcaption{Proposed method}\label{fig:BER_Curves_of_FRAC_FBMC_VA}
    \end{subfigure}  
    \hspace*{\fill}
\caption{BER curves for different schemes{ -- (a)-(c) Flat fading; (d)-(f) ITU-R-PA; (g)-(i) ITU-R-VA}}
    \label{fig:BER_Curves_for_diff_IAFO}
\end{figure}    
In this subsection, we compare the impact of IAFO on various existing Alamouti-coded FBMC schemes. 
In Fig. \ref{fig:BER_Curves_for_diff_IAFO}, the BER curves for TR-FBMC in \cite{Renfors 2010}, FBMC-QAM in \cite{Nam 2016}, and the proposed scheme, are plotted under flat channel (Fig. 3a-c), ITU-R-PA channel (Fig. 3d-f), and ITU-R-VA channel (Fig. 3g-i), respectively.
The half-subblock size $N_F/2$ for the proposed scheme in Figs. \ref{fig:BER_Curves_of_FRAC_FBMC_flat}, \ref{fig:BER_Curves_of_FRAC_FBMC_PA}, and \ref{fig:BER_Curves_of_FRAC_FBMC_VA} are set to 128, 8, and 4, respectively.
It is clearly observed that under all channels, the two existing FBMC schemes suffer BER degradation when the IAFO increases from 0 to 0.3 subcarrier spacing, as shown in Figs. 3ab, 3de, and 3gh, respectively.
This is because TR-FBMC and FBMC-QAM can not discard the interference of IAFO when we compensate for specific carrier frequency offset and thus it leads to the retention of inter-antenna ICI. 
However, as shown in Figs. \ref{fig:BER_Curves_of_FRAC_FBMC_flat}, \ref{fig:BER_Curves_of_FRAC_FBMC_PA}, and \ref{fig:BER_Curves_of_FRAC_FBMC_VA}, the proposed scheme only has negligible changes in BER with increasing IAFOs, demonstrating the proposed scheme's unprecedented advantages in the presence of a high IAFO.
    
\section{Conclusion}
In this letter, a new FRAC-FBMC scheme that has inter-antenna ICI-free characteristics even with nonzero IAFOs was proposed. 
By allowing the use of multiple subblocks, the proposed scheme works well even in frequency-selective channels, with only a slight sacrifice in bandwidth efficiency.
Simulation results confirmed the resilience of the proposed scheme to the IAFO and showed that even if the IAFO is as high as $0.3$ times subcarrier spacing, the proposed scheme outperforms other schemes for any frequency-selective channels. 

\ifCLASSOPTIONcaptionsoff
  \newpage
\fi


\begin{thebibliography}{1}

\bibitem{Renfors 2010}
M. Renfors, T. Ihalainen and T. H. Stitz, ``A block-Alamouti scheme for filter bank based multicarrier transmission,'' 2010 European Wireless Conference (EW), 2010, pp. 1031-1037.

\bibitem{Nam 2016}
H. Nam, M. Choi, S. Han, C. Kim, S. Choi and D. Hong, ``A New Filter-Bank Multicarrier System With Two Prototype Filters for QAM Symbols Transmission and Reception,'' in IEEE Transactions on Wireless Communications, vol. 15, no. 9, pp. 5998-6009, Sept. 2016.

\bibitem{Moerman 2022}
A. Moerman et al., ``Beyond 5G Without Obstacles: mmWave-over-Fiber Distributed Antenna Systems,'' in IEEE Communications Magazine, vol. 60, no. 1, January 2022, pp. 27-33.
\bibitem{Elhattab 2022}
M. Elhattab, M. A. Arfaoui and C. Assi, ``Joint Clustering and Power Allocation in Coordinated Multipoint Assisted C-NOMA Cellular Networks,'' in IEEE Transactions on Communications, vol. 70, no. 5, May 2022, pp. 3483-3498.

\bibitem{Na-Choi-2016}
D. Na and K. Choi, “Intrinsic ICI-free alamouti coded FBMC,” \emph{IEEE
Commun. Lett}., vol. 20, no. 10, pp. 1971–1974, Oct. 2016. 
\bibitem{Na-Choi-2017}
D. Na and K. Choi, ``Generalization of the phase shift condition
in ``Intrinsic ICI-free Alamouti coded FBMC'','' \emph{IEEE Commun. Lett}.,
vol. 21, no. 8, pp. 1747–1750, Aug. 2017.

\bibitem{Bellanger 2010}
M. Bellanger, ``FBMC physical layer: a primer'', PHYDYAS, Jan 2010.

\bibitem{ITU-R-M.1225}
\emph{Guidelines for Evaluation of Radio Transmission Technologies for
IMT-2000}, document ITU-R M.1225, 1997.

\end{thebibliography}
\end{document}